\documentclass[a4paper]{article}

\usepackage{amsmath,graphicx}
\usepackage{booktabs,arydshln}
\usepackage{INTERSPEECH2021}
\usepackage{xcolor}

\newcommand\blfootnote[1]{%
    \begingroup
    \renewcommand\thefootnote{}\footnote{#1}%
    \addtocounter{footnote}{-1}%
    \endgroup
}

\title{Comparing CTC and LFMMI for out-of-domain adaptation of wav2vec 2.0 acoustic model}

\name{Apoorv Vyas$^{\,1\,2}$, Srikanth Madikeri$^{\,1}$, Herv{\'e} Bourlard$^{\,1\,2}$}
\address{
    $^{1}$Idiap Research Institute, Martigny, Switzerland \\
    $^{2}$Ecole Polytechnique F\'ed\'erale de Lausanne, Switzerland
}
\email{\{avyas, msrikanth, bourlard\}@idiap.ch}

\begin{document}

\maketitle
\begin{abstract}
In this work, we investigate if the wav2vec 2.0 self-supervised pretraining
helps mitigate the overfitting issues with connectionist temporal classification (CTC) training
to reduce its performance gap with flat-start lattice-free MMI (E2E-LFMMI)
for automatic speech recognition with limited training data. Towards
that objective, we use the pretrained wav2vec 2.0 BASE model and fine-tune it
on three different datasets including out-of-domain (Switchboard) and
cross-lingual (Babel) scenarios.  Our results show that for supervised
adaptation of the wav2vec 2.0 model, both E2E-LFMMI and CTC achieve similar
results; significantly outperforming the baselines trained only with supervised
data. Fine-tuning the wav2vec 2.0 model with E2E-LFMMI and CTC we obtain the following relative WER improvements over the supervised baseline trained with E2E-LFMMI. We get relative improvements of $40$\% and $44$\% on the clean-set and $64$\% and $58$\% on the test set of Librispeech ($100$h) respectively. On Switchboard ($300$h) we obtain relative improvements of $33$\% and $35$\% respectively. Finally, for Babel languages, we obtain relative improvements of
$26$\% and $23$\% on Swahili ($38$h) and $18$\% and $17$\% on Tagalog ($84$h)
respectively.
\blfootnote{Preprint. Under Review.}

\end{abstract}


\section{Introduction}
\label{sec:intro}

Self-supervised training methods to learn powerful acoustic representations
from untranscribed audio data have received a lot of attention recently
\cite{baevski2020wav2vec,chung2019unsupervised,liu2020tera,oord2019representation,schneider2019wav2vec,wang2020speechbert}.
These methods aim to learn models that can extract good representations from
the audio signal. The learnt representations or the model can later be adapted
using supervised data to achieve state-of-the-art performance for automatic
speech recognition (ASR) while greatly reducing the amount of transcribed
training data which is both expensive and time-consuming to obtain. 

Self-supervised training methods can broadly be grouped into two categories:
(1) auto-regressive models that predict the future representations given only
the the past inputs \cite{chung2019unsupervised, oord2019representation} and
(2) bidirectional models that predict masked parts of the input by looking at
the full input \cite{baevski2020wav2vec,liu2020tera}. 

Currently, bidirectional models outperform autoregressive self-supervised
models for ASR \cite{baevski2020wav2vec,liu2020tera}.  In
\cite{baevski2020wav2vec}, the authors train a transformer model, wav2vec 2.0,
that learns representations from raw audio data using contrastive learning. The
model is trained on $1000$ hours of unsupervised Librispeech
\cite{panayotov2015librispeech} data and is later adapted on a $100$ hour
supervised subset of Librispeech data to achieve state-of-the-art performance. While wav2vec 2.0 model achieves state-of-the-art performance on Librispeech
$100$h subset, the authors only consider Connectionist Temporal Classification
(CTC) \cite{graves2006ctc} for acoustic model training. Moreover, they only
investigate the performance of supervised adaptation on subsets of Librispeech
dataset which was used for pretraining. 

In \cite{vyas2021lfmmi}, it was shown that self-supervised pretraining with
reconstruction based masked acoustic modeling \cite{liu2020tera} remains useful
on the out-of-domain datasets when the pretrained model is adapted with
flat-start lattice free MMI (E2E-LFMMI) \cite{hadian2018flat}. In this work, we
not only investigate the performance of wav2vec 2.0 pretraining in such
scenarios but we also look into the role of the training criterion for
supervised adaptation of wav2vec 2.0 model.

The contribution of this work is twofold. First, we compare the effect of
sequence discriminative training criterion for supervised adaptation. We show
that fine-tuning the wav2vec 2.0 model with E2E-LFMMI and CTC criterion yield
similar performances with neither consistently better than the other. Second,
we adapt the wav2vec 2.0 model on out-of-domain conversational speech and on
cross-lingual data to show that the wav2vec 2.0 pretraining provides
significant gains over the models trained only with supervised data. 

Specifically, we fine-tune the wav2vec 2.0 BASE model \cite{baevski2020wav2vec}
pretrained on $1000$ hours of Librispeech untranscribed data on three
different datasets using E2E-LFMMI and CTC criteria. We first consider a
$100$h subset of Librispeech which was also seen during the pretraining
stage. We next train on the three hundred hours of Switchboard data, which is
conversational telephonic speech sampled at $8$ KHz as opposed to the
pretraining data which is read speech sampled at $16$ KHz. Finally, we also
evaluate on two low resource languages, Tagalog and Swahili, from the Babel
dataset. To the best of our knowledge, we are the first ones to show that both
E2E-LFMMI and CTC training achieve similar results on low resource languages
when fine-tuned with a pretrained model.

The rest of the paper is organized as follows: In Section \ref{sec:method}, we
describe the details such as model architecture, fine-tuning learning rate, and
decoding hyper-parameters. In Section \ref{sec:asr}, we present the details of
the data preparation and baselines we considered. Finally, in Section
\ref{subsec:results}, we present the results on comparison between E2E-LFMMI
and CTC on Librispeech, Switchboard, and Babel datasets.
\section{Our Method}
\label{sec:method}
To investigate the effectiveness of wav2vec 2.0 \cite{baevski2020wav2vec}
pretraining for cross-lingual transfer and out of domain adaptation, we choose
the BASE model as our pretrained model. This model is pretrained on $1000$
hours of Librispeech data. The model contains $12$ transformer layers each with
$8$ attention heads. The embedding dimension is set to $96$ for each head and
feed-forward dimension is set to $3072$. In the following subsections we
provide  details of acoustic model training and decoding.

\subsection{Acoustic Model Training}
The input to the wav2vec 2.0 BASE model is the raw speech signal sampled at
$16$ KHz. During supervised adaptation, we pass the output of the wav2vec 2.0
transformer encoder to a seven layered factorized time-delay neural network
(TDNNF). We fine-tune the wav2vec 2.0 BASE model together with TDNNF layers
using E2E-LFMMI and CTC criteria.

Our E2E-LFMMI model is trained with biphone units while our CTC model is
trained using character units. We train for a maximum of $30000$ and $75000$
updates for E2E-LFMMI and CTC criterion respectively where each update is over
$1500$ seconds of speech input. All our models are trained with $3$ GTX-$1080$
Ti GPUs with gradient synchronized training. We use gradient accumulation to
obtain a batch size of $1500$ seconds.

For both E2E-LFMMI and CTC, we update the BASE model parameters with a
learning rate that is linearly increased to 3e-5 over $10$\% of the updates,
then held constant for $40$\% of the updates, and linearly decreased for $50$\%
of the updates. For TDNNF model parameters, we use a learning rate that is $20$
times the current learning rate for BASE model updates. We also use the natural
gradient update \cite{povey2014parallel} for training with the E2E-LFMMI
objective. 

All our models are trained with PyTorch \cite{paszke2019pytorch} \footnote{We
will open-source our training scripts and models}. For fine-tuning with CTC, we
use the Fairseq toolkit \cite{ott2019fairseq} and for E2E-LFMMI, we use the
Espresso toolkit \cite{wang2019espresso} which uses PyChain
\cite{shao2020pychain} for the implementation of LFMMI loss. We use the PyTorch
implementation for natural gradient update from \cite{madikeri2020pkwrap}.

\subsection{Decoding}
For all models trained with E2E-LFMMI, we use the WFST decoder from
\cite{povey2011kaldi} with a beam width of $15$. For the models trained with
CTC, we use the decoder from \cite{pratap2019wav2letter} with a beam width of
$500$. We always use the language model from Kaldi recipes
\cite{povey2011kaldi} which are trained with SRILM \cite{stolcke2002srilm}. We
found this to give better results than KenLM \cite{heafield2011kenlm}.

\section{Experiments}
\label{sec:asr}
In the following, we discuss in detail the datasets and augmentation that we
used, baselines we compare against, and finally the results for the
experiments.

\subsection{Datasets}
\label{subsec:datasets}
We evaluate the performance of the wav2vec 2.0 BASE model on the same datasets as \cite{vyas2021lfmmi}. The three datasets are selected in increasing order of difficulties.  We first consider the $100$ hour clean
subset of Librispeech \cite{panayotov2015librispeech}. This is the easiest
setting because the training set is a part of the $1000$ hours of Librispeech
pretraining data. We next consider the Switchboard
\cite{godfrey1992switchboard} dataset with $300$ hours of transcribed data. In
contrast to Librispeech pretraining data, Switchboard has conversational speech
sampled at $8$ KHz making it a more difficult out-of-domain setting. We finally
fine-tune on two of the Babel \cite{gales2014speech} languages: Tagalog ($84$h)
and Swahili ($38.5$h). We consider this to be the hardest setting as there is both
language and acoustic conditions mismatch. For all our experiments we apply
speed and volume perturbation to increase the dataset by three times. 

\subsection{Baselines}
\label{subsec:baselines}
We reproduce the supervised baselines from \cite{vyas2021lfmmi}. We train from
scratch a twelve layered TDNNF model using $80$ dimensional filter bank features.
Our baseline is trained with the E2E-LFMMI objective. For all TDNNF models, we
set hidden layer dimension to $1024$ and bottleneck dimension to $128$. 

We also compare against pretrained model trained with Masked Acoustic Modeling
(MAM) objective \cite{liu2020tera}. For the MAM pretraining, the input to the
network is a sequence of masked or noise corrupted acoustic features. The model
attempts to reconstruct the original input given the corrupted input and is
trained with $\text{L}_1$ loss. This is different from wav2vec 2.0 pretraining
which masks input segments and uses cross-entropy loss to contrast between
learned representation at the masked time-steps with representations at other
time steps.  

For comparison against MAM pretraining, we refer to results from
\cite{vyas2021lfmmi} where the pretrained model was fine-tuned with E2E-LFMMI
objective on the same datasets. Note that this model had $6$ attention heads
per layer in comparison to wav2vec 2.0 base model which has $8$ attention
heads per layer. This results in a model of smaller capacity. We believe that
this doesn't affect our conclusions as there is no significant performance
difference between models of different capacity that are pretrained with MAM
\cite{liu2020tera}. 

Finally, when available, we provide results for models trained with CTC
criterion using only the supervised data. Note that the dataset augmentation
and model architectures can differ significantly from our TDNNF baseline trained with E2E-LFMMI criterion and 
they might not be strictly comparable.

\subsection{Results}
\label{subsec:results}
In the following we compare the Word Error Rate (WER) achieved by supervised
adaptation of pretrained wav2vec 2.0 BASE model using E2E-LFMMI and CTC
objectives. We refer to the $12$ layered TDNNF baseline trained only on
supervised data as \emph{TDNNF}. Our wav2vec 2.0 BASE model with seven TDNNF layers is referred to as \emph{wav2vec2-base}. The masked acoustic pretrained model
 from \cite{vyas2021lfmmi} is referred to as \emph{MAM}.

\subsubsection{Librispeech ($100$ hours)}
\label{subsubsec:libri}

In this experiment, we discuss the case when the supervised training data was
seen during the pretraining. We present our main results in Table
\ref{tab:librispeech}. For evaluation, we select the model that achieves lowest WER
on the \emph{dev-other} set. For E2E-LFMMI models, we first decode using a
$3$-gram language model and then rescore using a $4$-gram language model. For
the CTC model, we directly decode with the $4$-gram language model as done in
\cite{baevski2020wav2vec}. The beam search decoder from
\cite{pratap2019wav2letter} can efficiently decode with $4$-gram language model on a
single GPU.

In rows (a) and (b), we present the results for training with only supervised
data using E2E-LFMMI and CTC. From the comparison on the \emph{dev} set it can
be seen that CTC training requires additional regularization and modifications
to the deep neural network training to reach a similar level of performance.

In rows (d) and (e), we compare the performance of fine-tuning the wav2vec 2.0
BASE model with E2E-LFMMI and CTC loss. It can be seen that both models
reach similar level of performance providing $\sim12.7$\% and $\sim11.5$\% absolute
WER improvements over the supervised \emph{TDNNF} baseline on the noisy portion of
the test set. Note that we did not apply any additional regularization or
changes to train with the CTC loss. 

\begin{table*}[ht!]
    \centering
    \begin{tabular}{clccccccccccccc}
    \toprule
          &                                   &             & \phantom{a} & \textbf{dev}    &  \phantom{a} & \multicolumn{5}{c}{\textbf{test}} \\
          &                                   &             && \textbf{3-gram} && \multicolumn{2}{c}{\textbf{3-gram}} & \phantom{a} &\multicolumn{2}{c}{\textbf{4-gram}} \\
          & Architecture                      & Criterion   && clean  && clean    & other    && clean    & other \\
      
      \midrule
      \multicolumn{14}{c}{Supervised Only} \\
      \midrule
      (a) & TDNNF-large                       & E2E-LFMMI   && 8.3    && 8.6      & 26.3     && 6.0      & 19.9 \\
      (b) & Bi-LSTM \cite{billa2017improving} & CTC         && 11.1   && -        & -        && -        & - \\
       & \quad + max perturbation             & CTC         && 9.8    && -        & -        && -        & - \\
       & \quad + cascade dropout              & CTC         && 7.9    && 8.7      & 26.1     && -        & - \\

      \midrule
      \multicolumn{14}{c}{Pretraining + Supervised (others)} \\
      \midrule
      (c) & MAM \cite{vyas2021lfmmi}          & E2E-LFMMI   && -      && 7.8      & 20.2     && 5.3      & 14.7 \\
      
      \midrule
      \multicolumn{14}{c}{Pretraining + Supervised (ours)} \\
      \midrule
      (d) & wav2vec2-base                          & E2E-LFMMI   && -      && \bf{4.4} & \bf{8.9} && 3.5      & \bf{7.3} \\
      (e) & wav2vec2-base                          & CTC         && -      && -        & -        && \bf{3.3} & 8.5 \\

       \bottomrule
    \end{tabular}
    \vspace{4pt}
    \caption{
        Comparison of word error rates (WER) (in \%) on the clean and other
        parts of the Librispeech test set with and without 4-gram language
        model rescoring. Fine-tuning the pretrained wav2vec 2.0 BASE model with
        E2E-LFMMI or CTC gives similar level of performance without any
        additional regularization techniques. Both models outperform the
        masked-acoustic modeling (MAM) pretraining and fully supervised
        baselines.
    }
    \label{tab:librispeech}
\end{table*}

Furthermore it can be noticed from (c) that wav2vec 2.0 BASE model fine-tuned
with either loss significantly outperforms the model pretrained with masked
acoustic modeling.

\subsubsection{Switchboard ($300$ hours)}
\label{subsubsec:swbd}

In this experiment, we explore the out-of-domain scenario in which pretraining
data and supervision data share the same language however they are dissimilar
with respect to content, and acoustic conditions. Switchboard dataset comprises
of telephonic conversations which are recorded at $8$ KHz. This is different
from Librispeech pretraining data which comprises of read speech sampled at
$16$ KHz. 

For fine-tuning with wav2vec 2.0 BASE model, we resample the Switchboard
recordings at $16$ KHz. To train the baseline \emph{TDNNF} acoustic model using
only the transcribed data, we use the $8$ KHz recordings.

For evaluation we select the model that gives smallest WER on the held out
development set. Once again, for E2E-LFMMI models, we first decode using a
$3$-gram language model followed by rescoring with $4$-gram language model
trained on Switchboard and Fisher transcripts. For model fine-tuned with CTC,
we directly decode with the $4$-gram language model.

Table \ref{tab:swbd} compares the WER for the E2E-LFMMI and CTC  models trained
from scratch as well as those fine-tuned from models pretrained on Librispeech
data. Consistent with Librispeech experiment, we see that for models trained using only the supervised data, CTC requires
additional regularization techniques to reach the same level of performance as
E2E-LFMMI. Furthermore, the CTC baseline presented in
\cite{audhkhasi2019forget} applies fMLLR transformation which typically
provides additional gains for end-to-end ASR \cite{tomashenko2018evaluation}.

Once again, it can be seen that fine-tuning wav2vec 2.0 BASE model with
E2E-LFMMI or CTC leads to comparable performances. Both models significantly
outperform the baselines trained only with supervised data as well as the model
pretrained with masked acoustic modeling. Similar to the previous experiment,
we do not use any additional regularization techniques for CTC training. Note
that we get absolute WER improvements of $\sim3.7$\% and $\sim7$\% over the \emph{TDNNF}
baseline on the switchboard and callhome portion of evaluation sets.

\begin{table}[th]
    \centering
    \scalebox{0.9}{
        \begin{tabular}{lccccc}
        \toprule
        & & \multicolumn{4}{c}{\textbf{Hub5'00 (eval2000)}} \\
        & & \multicolumn{2}{c}{\textbf{3-gram}} & \multicolumn{2}{c}{\textbf{4-gram}} \\
        \textbf{Model}                     & \textbf{Criterion} &\textbf{SW} & \textbf{CH} & \textbf{SW} & \textbf{CH} \\
        \midrule
        \multicolumn{6}{c}{Supervised Only} \\
        \midrule
        TDNNF-large                        & E2E-LFMMI          & 11.8       & 22.5        & 10.3        & 20.3 \\
        Bi-LSTM \cite{audhkhasi2019forget} & CTC                & -          & -           & 12.2        & 21.8 \\
        \quad + seq. noise                 & CTC                & -          & -           & 10.9        & 21.3 \\  
        \quad + regularization            & CTC                & -          & -           & 10.6        & 19.5 \\
        \midrule
        \multicolumn{6}{c}{Pretraining + Supervised (Others)} \\
        \midrule
        MAM \cite{vyas2021lfmmi}           &  E2E-LFMMI         & 10.9       & {20.4}      & {9.4}       & {18.2} \\
         \midrule
         \multicolumn{6}{c}{Pretraining + Supervised (Ours)} \\
         \midrule
        wav2vec2-base                           & E2E-LFMMI          & \bf{7.3}   & \bf{14.5}   & 6.7         & 13.7 \\
        wav2vec2-base                           & CTC                & -          & -           & \bf{6.6}    & \bf{13.2} \\
        \bottomrule
        \end{tabular}
    }
    \vspace{4pt}
    \caption{
        Comparison of word error rates (WER) (in \%) on eval2000 test set for
        the 300 hours Switchboard task. The 3-gram language model is based on
        Switchboard, whereas the 4-gram employs Switchboard+Fisher training set
        transcripts. Fine-tuning wav2vec 2.0 BASE model with E2E-LFMMI or CTC
        improves the performance over the baselines with no additional
        regularization needed for the CTC model. 
    }
    \label{tab:swbd}
\end{table}

\subsubsection{Babel: Swahili and Tagalog}
\label{subsubsec:babel}
In this experiment, we evaluate the effectiveness of wav2vec 2.0 pretraining
when the BASE model is fine-tuned on cross lingual  data. For our
evaluation, we consider two low resource languages, Swahili
and Tagalog, from the Babel dataset. Once again, we resample the audio recordings at $16$ KHz for
fine-tuning the pretrained model. For our \emph{TDNNF} baseline trained only
supervised data, we use the original recordings sampled at $8$ KHz.

For both languages, we report the results on the \emph{dev10h} development part
due to the lack of a separate evaluation set. We use the $2$ hour development
set for model selection. For both  E2E-LFMMI and CTC models, we use $3$-gram
language model for decoding using the previously described hyperparameters for
beam search. We do not consider the non-language symbols for scoring on these
datasets.

Table \ref{tab:babel} compares the WER for the models trained from
scratch to the models pretrained on the Librispeech dataset. Note that we could
not find any CTC baseline that is trained only on the supervised data and
provides competitive performance to E2E-LFMMI training. 

Once again, we see that for both Swahili and Tagalog, wav2vec 2.0 BASE model
fine-tuned with E2E-LFMMI and CTC obtain similar performance. Both models
outperform the baselines trained only with supervised data as well as the model 
pretrained with masked acoustic modeling.

We additionally report the WER for the XLSR-$10$ model from
\cite{conneau2020unsupervised}. This is a large wav2vec 2.0 multilingual model
pretrained on $10$ languages. As can be seen, we get a much better word error
rate on Swahili and a comparable performance on Tagalog. We think that the
results might not be directly comparable because we use speed and volume perturbation for data augmentation and do not score on non-language
symbols. Additionally, XLSR-$10$ uses KenLM for decoding while we use SRILM. In
our experiments, we noticed a significant degradation in WER using KenLM.
Despite these differences, it is clear from our results that wav2vec 2.0
BASE model pretrained on Librispeech still offers a very competitive baseline
to the large multilingual model.

\begin{table}[th]
    \centering
    \scalebox{0.9}{
    \begin{tabular}{cccc}
        \toprule
        \textbf{Model}                                  & \textbf{Criterion} & \textbf{Swahili} & \textbf{Tagalog} \\
        \midrule
         \multicolumn{4}{c}{{Supervised Only}} \\
         \midrule
         TDNNF   & E2E-LFMMI  & 39.5     & 44.9 \\
         \midrule
         \multicolumn{4}{c}{Pretraining + Supervised (Others)} \\
         \midrule
         MAM \cite{vyas2021lfmmi}                       & E2E-LFMMI          & 36.7             & 43.4  \\
         XLSR-10 (Large) \cite{conneau2020unsupervised} & CTC                & 35.5             & 37.3 \\
         \midrule
         \multicolumn{4}{c}{{Pretraining + Supervised} (Ours)} \\
         \midrule
         wav2vec2-base                                       & E2E-LFMMI          & \bf{29.4}        & \bf{36.9}  \\
         wav2vec2-base                                       & CTC                & 30.4             & 37.3  \\
        \bottomrule
    \end{tabular}
    }
    \vspace{4pt}
    \caption{
        Comparison of word error rates (WER) (in \%) on dev10h set for
        the Swahili and Tagalog languages of the Babel dataset. Fine-tuning
        the pretrained wav2vec 2.0 BASE model significantly outperforms the
        monolingual and MAM baselines. Note that while we use SRILM, XLSR-10 model uses KenLM for decoding and does not use speed or volume perturbation. 
    }
    \label{tab:babel}
\end{table}

\section{Conclusions}
In this work, we investigate the effects of the sequence discriminative
training criteria for the supervised adaptation of pretrained wav2vec 2.0 BASE
model.  We show that fine-tuning wav2vec 2.0 BASE model with either E2E-LFMMI
or CTC gives similar performance with no additional regularization needed for
CTC training. We further show that wav2vec 2.0 pretraining provides significant
gains and outperforms models pretrained with masked acoustic modeling even for
out-of-domain and cross-lingual adaptation. 

In future, we will compare the performance of the monolingual wav2vec 2.0 model
to the multilingual model of similar capacity to understand the advantages of
multilingual pretraining. We will additionally explore supervised fine-tuning
with multilingual data to further improve the performance in low resource
settings.

\section{Acknowledgments}
The research leading to these results has received funding from
Swiss National Science Foundation project SHISSM (Sparse and
hierarchical Structures for Speech Modeling), grant agreement 200021-175589. 
The research is also partially based upon the work supported by the Office of
the Director of National Intelligence (ODNI), Intelligence Advanced Research
Projects Activity (IARPA), via AFRL Contract \#FA8650-17-C-9116. The views and
conclusions contained herein are those of the authors and should not be
interpreted as necessarily representing the official policies or endorsements,
either expressed or implied, of the ODNI, IARPA, or the U.S. Government. The
U.S. Government is authorized to reproduce and distribute reprints for
Governmental purposes notwithstanding any copyright annotation thereon.
\bibliographystyle{IEEEtran}

\bibliography{mybib}

\begin{thebibliography}{10}
\providecommand{\url}[1]{#1}
\csname url@samestyle\endcsname
\providecommand{\newblock}{\relax}
\providecommand{\bibinfo}[2]{#2}
\providecommand{\BIBentrySTDinterwordspacing}{\spaceskip=0pt\relax}
\providecommand{\BIBentryALTinterwordstretchfactor}{4}
\providecommand{\BIBentryALTinterwordspacing}{\spaceskip=\fontdimen2\font plus
\BIBentryALTinterwordstretchfactor\fontdimen3\font minus
  \fontdimen4\font\relax}
\providecommand{\BIBforeignlanguage}[2]{{%
\expandafter\ifx\csname l@#1\endcsname\relax
\typeout{** WARNING: IEEEtran.bst: No hyphenation pattern has been}%
\typeout{** loaded for the language `#1'. Using the pattern for}%
\typeout{** the default language instead.}%
\else
\language=\csname l@#1\endcsname
\fi
#2}}
\providecommand{\BIBdecl}{\relax}
\BIBdecl

\bibitem{baevski2020wav2vec}
A.~Baevski, H.~Zhou, A.~Mohamed, and M.~Auli, ``wav2vec 2.0: A framework for
  self-supervised learning of speech representations,'' in \emph{Proceedings of
  the international conference on Neural Information Processing Systems
  (NeurIPS)}, 2020.

\bibitem{chung2019unsupervised}
Y.-A. Chung, W.-N. Hsu, H.~Tang, and J.~Glass, ``An unsupervised autoregressive
  model for speech representation learning,'' in \emph{Interspeech}, 2019.

\bibitem{liu2020tera}
A.~T. Liu, S.-W. Li, and H.~yi~Lee, ``Tera: Self-supervised learning of
  transformer encoder representation for speech,'' \emph{arXiv}, 2020.

\bibitem{oord2019representation}
A.~van~den Oord, Y.~Li, and O.~Vinyals, ``Representation learning with
  contrastive predictive coding,'' \emph{arXiv}, 2019.

\bibitem{schneider2019wav2vec}
S.~Schneider, A.~Baevski, R.~Collobert, and M.~Auli, ``wav2vec: Unsupervised
  pre-training for speech recognition,'' in \emph{Interspeech}, 2019.

\bibitem{wang2020speechbert}
W.~Wang, Q.~Tang, and K.~Livescu, ``Unsupervised pre-training of bidirectional
  speech encoders via masked reconstruction,'' in \emph{Proceedings of ICASSP},
  2020.

\bibitem{panayotov2015librispeech}
V.~{Panayotov}, G.~{Chen}, D.~{Povey}, and S.~{Khudanpur}, ``Librispeech: An
  asr corpus based on public domain audio books,'' in \emph{Proceedings of
  ICASSP}, 2015.

\bibitem{graves2006ctc}
A.~Graves, S.~Fern\'{a}ndez, F.~Gomez, and J.~Schmidhuber, ``Connectionist
  temporal classification: Labelling unsegmented sequence data with recurrent
  neural networks,'' in \emph{Proceedings of the 23rd International Conference
  on Machine Learning}, 2006.

\bibitem{vyas2021lfmmi}
A.~Vyas, S.~Madikeri, and H.~Bourlard, ``Lattice-free mmi adaptation of
  self-supervised pretrained acoustic models,'' in \emph{Proceedings of
  ICASSP}, 2021.

\bibitem{hadian2018flat}
H.~Hadian \emph{et~al.}, ``Flat-start single-stage discriminatively trained
  hmm-based models for asr,'' \emph{IEEE ACM Transactions on Audio, Speech, and
  Language Processing}, 2018.

\bibitem{povey2014parallel}
D.~{Povey}, X.~{Zhang}, and S.~{Khudanpur}, ``{Parallel training of DNNs with
  Natural Gradient and Parameter Averaging},'' \emph{arXiv}, 2014.

\bibitem{paszke2019pytorch}
A.~Paszke, S.~Gross, F.~Massa, A.~Lerer, J.~Bradbury, G.~Chanan, T.~Killeen,
  Z.~Lin, N.~Gimelshein, L.~Antiga, A.~Desmaison, A.~Kopf, E.~Yang, Z.~DeVito,
  M.~Raison, A.~Tejani, S.~Chilamkurthy, B.~Steiner, L.~Fang, J.~Bai, and
  S.~Chintala, ``Pytorch: An imperative style, high-performance deep learning
  library,'' in \emph{Advances in Neural Information Processing Systems 32},
  2019.

\bibitem{ott2019fairseq}
M.~Ott, S.~Edunov, A.~Baevski, A.~Fan, S.~Gross, N.~Ng, D.~Grangier, and
  M.~Auli, ``fairseq: A fast, extensible toolkit for sequence modeling,'' in
  \emph{Proceedings of NAACL-HLT 2019: Demonstrations}, 2019.

\bibitem{wang2019espresso}
Y.~Wang, T.~Chen, H.~Xu, S.~Ding, H.~Lv, Y.~Shao, N.~Peng, L.~Xie, S.~Watanabe,
  and S.~Khudanpur, ``Espresso: A fast end-to-end neural speech recognition
  toolkit,'' in \emph{IEEE Automatic Speech Recognition and Understanding
  Workshop (ASRU)}, 2019.

\bibitem{shao2020pychain}
Y.~Shao, Y.~Wang, D.~Povey, and S.~Khudanpur, ``{ PyChain: A Fully Parallelized
  PyTorch Implementation of LF-MMI for End-to-End ASR},'' in \emph{Proc.
  Interspeech}, 2020.

\bibitem{madikeri2020pkwrap}
S.~Madikeri, S.~Tong, J.~Zuluaga-Gomez, A.~Vyas, P.~Motlicek, and H.~Bourlard,
  ``Pkwrap: a pytorch package for lf-mmi training of acoustic models,''
  \emph{arXiv}, 2020.

\bibitem{povey2011kaldi}
D.~Povey, A.~Ghoshal, G.~Boulianne, L.~Burget, O.~Glembek, N.~Goel,
  M.~Hannemann, P.~Motlicek, Y.~Qian, P.~Schwarz, J.~Silovsky, G.~Stemmer, and
  K.~Vesely, ``The kaldi speech recognition toolkit,'' in \emph{IEEE 2011
  Workshop on Automatic Speech Recognition and Understanding}, 2011.

\bibitem{pratap2019wav2letter}
V.~{Pratap}, A.~{Hannun}, Q.~{Xu}, J.~{Cai}, J.~{Kahn}, G.~{Synnaeve},
  V.~{Liptchinsky}, and R.~{Collobert}, ``Wav2letter++: A fast open-source
  speech recognition system,'' in \emph{ICASSP 2019 - 2019 IEEE International
  Conference on Acoustics, Speech and Signal Processing (ICASSP)}, 2019.

\bibitem{stolcke2002srilm}
A.~Stolcke, ``Srilm - an extensible language modeling toolkit.'' in
  \emph{Proceedings of Interspeech}, 2002.

\bibitem{heafield2011kenlm}
K.~Heafield, ``{K}en{LM}: Faster and smaller language model queries,'' in
  \emph{Proceedings of the Sixth Workshop on Statistical Machine Translation},
  2011.

\bibitem{godfrey1992switchboard}
J.~J. Godfrey, E.~C. Holliman, and J.~McDaniel, ``Switchboard: Telephone speech
  corpus for research and development,'' in \emph{IEEE International Conference
  on Acoustics, Speech, and Signal Processing,}, 1992.

\bibitem{gales2014speech}
M.~J.~F. Gales, K.~M. Knill, A.~Ragni, and S.~P. Rath, ``Speech recognition and
  keyword spotting for low-resource languages: Babel project research at
  cued.'' in \emph{The Workshop on Spoken Language Technologies for
  Under-Resourced Languages}, 2014.

\bibitem{billa2017improving}
J.~Billa, ``Improving lstm-ctc based asr performance in domains with limited
  training data,'' \emph{arXiv}, 2017.

\bibitem{audhkhasi2019forget}
K.~Audhkhasi, G.~Saon, Z.~Tüske, B.~Kingsbury, and M.~Picheny, ``{Forget a Bit
  to Learn Better: Soft Forgetting for CTC-Based Automatic Speech
  Recognition},'' in \emph{Proc. Interspeech 2019}, 2019.

\bibitem{tomashenko2018evaluation}
N.~Tomashenko and Y.~Est{\`e}ve, ``{Evaluation of Feature-Space Speaker
  Adaptation for End-to-End Acoustic Models},'' in \emph{{LREC 2018}}, 2018.

\bibitem{conneau2020unsupervised}
A.~Conneau, A.~Baevski, R.~Collobert, A.~Mohamed, and M.~Auli, ``Unsupervised
  cross-lingual representation learning for speech recognition,'' \emph{arXiv},
  2020.

\end{thebibliography}

\end{document}